# Swept-source heterodyne detection for interferometry in the H band

Félix Gudin*[a], Nicolas Forget [a]

[a]Université côte d'azur, CNRS, Institut de Physique de Nice, 17 rue Julien Lauprêtre, 06200 Nice, France

## ABSTRACT

It is well established that direct-detection interferometry outperforms heterodyne interferometry in terms of signal-to-noise ratio at optical wavelengths, not only because of the limited electronic bandwidth of photodetectors (typically <100 GHz), but also due to the shot-noise contribution of the optical local oscillator. The large-scale production of high-quality and cost-effective telecom components in the J+H band, and more recently in the K band, may open new schemes for the heterodyne signal detection and processing (e.g. correlation) as well as for the transport, distribution and control of the local oscillator. We describe how high-repetition-rate frequency-swept sources and telecom components could be exploited to parallelize, at reasonable cost, a large number of coherent detectors to expand the spectral coverage. We also present a first laboratory single-channel proof of concept that achieves a signal-to-noise ratio (SNR) probably sufficient for observing the brightest stars in the H band with telescopes of ~1 $m^2$ at short integration time (<100 ms).

## 1. INTRODUCTION

Long-baseline interferometry drives ever-finer angular resolution, from VLTI and CHARA [1,2] to the promising revival of intensity interferometry [3]. In this resolution pursuit, optical heterodyne interferometry is a compelling approach for combining signals from widely separated telescopes, avoiding the challenges of long-distance transport and stringent optical-path stabilization required by direct interferometry [4]. Pioneering work by Charles Townes and collaborators in the thermal infrared [5] demonstrated the power of this technique, achieving extreme angular resolutions on the order of a few tens of milliarcseconds (≈ 0.2 µrad) at 10.6 µm. However, extending this approach to optical wavelengths faces a significant bottleneck: an inherently low signal-to-noise ratio (SNR) [6], limited by (1) the narrow detection bandwidth of electronic detectors (<100 GHz) with regards to the bandwidth of thermal sources and (2) the shot-noise fluctuations of the single-frequency reference laser.

We introduce a novel detection concept that overcomes the fundamental bandwidth limitation and enables spectrally multiplexed detection. Rather than relying on a single-frequency laser, we employ a chirped mode-locked femtosecond laser whose wavelength-to-time mapping provides a frequency-swept local oscillator. The proposed architecture also relies on mature telecommunication-grade fiber-optic and photonic detection technologies, providing a scalable, commercially available, and cost-effective route for parallel multichannel detection. This paper presents the concept and an experimental proof-of-concept of this new heterodyne detection scheme.

## 2. MULTIPLEXING BROADBAND HETERODYNE DETECTION

In optical heterodyne interferometry, a beatnote between the faint astronomical light collected at the telescope focus and a strong local oscillator (LO) is recorded by high-speed photodetectors Fig 1. (a). The resulting frequency down conversion, arising from the frequency difference between the local oscillator and the signal within the photodetector bandwidth, enables RF transport and correlation to compute fringe visibility and retrieve source information [7].

*Contact author: felix.gudin@univ-cotedazur.fr

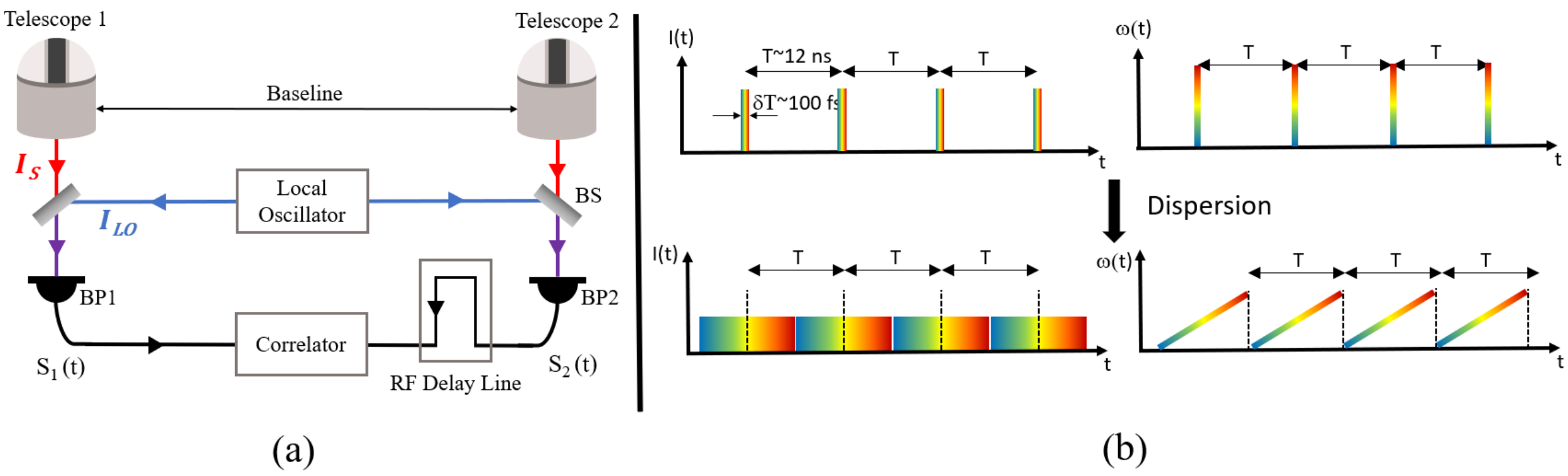

**Figure 1. (a)** Heterodyne optical interferometer architecture with a common local oscillator. **(b)** Time–frequency representation of a stretched pulse train. When the pulse duration matches the inter-pulse interval, the train forms a quasi-continuous swept-source local oscillator.

In this work, in contrast to conventional heterodyne interferometry, the local oscillator (LO) is not a low-noise, single-frequency continuous-wave laser but a train of ultrashort pulses from a mode-locked femtosecond laser that has been temporally stretched in a dispersive fiber [8]. As illustrated in Fig. 1(b) in both the time domain and the time–frequency plane, the dispersion is adjusted so that each pulse is stretched to match the pulse repetition period (12 ns in the experiments presented below). The resulting LO therefore behaves as a quasi-continuous source consisting of a train of broadband chirped pulses whose instantaneous optical frequency sweeps linearly with time. This periodic frequency sweep extends the spectral coverage accessible to heterodyne detection without requiring any increase in photoreceiver bandwidth. As the swept source undergoes additional dispersion, or as its optical bandwidth is increased, the stretched pulses eventually overlap in time. Consequently, frequency components originating from successive pulses become simultaneously present, allowing a wavelength-division multiplexing (WDM) demultiplexer to partition the broadband LO into *N* independent spectral channels. Each channel then carries a narrowband, continuously swept local oscillator, enabling parallel coherent detection. By leveraging commercially available low-loss coarse WDM (CWDM) components, this architecture can be scaled to several tens of spectral channels, substantially extending the accessible spectral coverage and/or increasing the single-to-noise ratio of the spectrally integrated detection. Furthermore, the intrinsic periodicity of the local oscillator provides a stable timing reference, eliminating the need for active stabilization or an external frequency reference. Last, the proposed concept offers an alternative to frequency-comb-based approaches by avoiding the need to align the line spacing and the individual comb lines with the channel spacing and bandwidth constraints imposed by WDM demultiplexing.

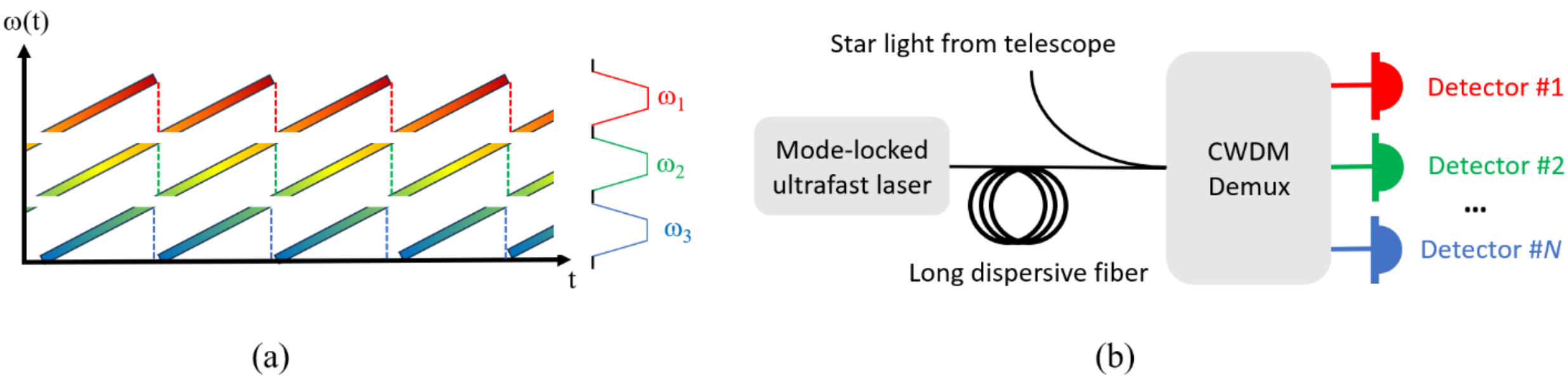

**Figure 2. (a)** Time–frequency representation of temporally overlapping pulses spectrally demultiplexed using a CWDM to generate multiple channels of swept-source local oscillators. **(b)** Schematic of the parallel heterodyne detection scheme using a commercial unstabilized mode-locked laser and standard fiber-optic components.

## 3. EXPERIMENTAL SINGLE-CHANNEL PROOF OF CONCEPT

We demonstrate the feasibility of this concept using a two-arm, single-spectral-channel, polarization-maintaining heterodyne interferometer built entirely from cost-effective, all-telecom-fiber components operating in the H band (~1.56 µm). The short pulses of the local oscillator are spectrally filtered and temporally stretched by a long dispersive fiber until their pulse duration matches the inter-pulse time interval (12 ns). To prevent pulse overlap and ensure single-channel operation, an 8 nm bandpass filter (BPF) centred at 1560 nm is inserted in the LO path. An identical BPF is placed in the signal arm, where a super luminescent diode (SLED) emulates stellar emission, ensuring matched optical bandwidths in the two arms.

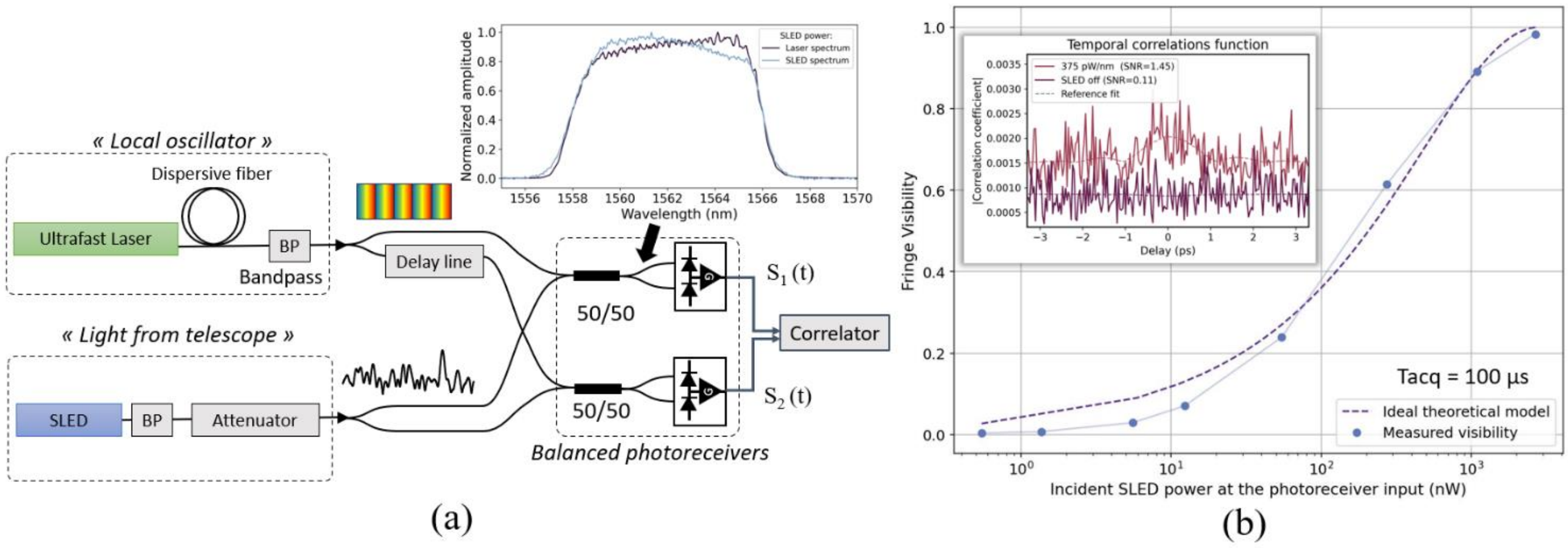

**Figure 3. (a)** All-polarization-maintaining fiber interferometer setup. Normalized spectra of the SLED and LO at the balanced photoreceiver input. Delay line: motorized free-space stage with 40 cm travel range and 50 nm precision; BPF: 8 nm bandpass filter; SLED: super luminescent diode; Correlator: 33 GHz, 100 GS/s oscilloscope. (**b)** Evolution of the fringe visibility as a function of SLED power, used to assess the sensitivity of the swept-source heterodyne detection system. The inset shows the correlation peak measured at the system sensitivity limit, corresponding to a SLED spectral power density of 375 pW/nm**.**

Beat notes between the LO and the signal are recorded using high-speed balanced photodetectors. Each detector comprises a pair of matched 22 GHz photodiodes followed by a differential amplifier that subtracts the photocurrents, thereby suppressing common-mode noise, particularly the laser relative intensity noise (RIN), while preserving the heterodyne signal. This balanced detection scheme enables operation close to the shot-noise limit. Fringe visibility is then retrieved via direct numerical correlation of the beat notes recorded on a 33 GHz oscilloscope. By varying the delay $\tau$ between the two arms, we retrieve the temporal coherence function of the artificial star, which corresponds to the Fourier transform of the source's spectral distribution, in practical terms, its spectrum. At each delay, we compute the fringe visibility, mathematically defined as the modulus of the Pearson correlation coefficient:

$$\rho_{1,2}(\tau) = \frac{\langle S_1(t) S_2(t-\tau)\rangle}{\sqrt{\langle S_1^2(t)\rangle\langle S_2^2(t-\tau)\rangle}} \tag{1}$$

To assess the sensitivity of the swept-source heterodyne detection system, we progressively reduce the input power until the correlation peak becomes indistinguishable from the noise floor. Despite residual LO noise arising from imperfect balanced detection and statistical photoreceiver amplifier noise (which prevent the correlation coefficient from reaching unity) an SNR > 1 is achieved with an input spectral density below 375 pW/nm at 1560 nm for an integration time of 0.4 ms. Extrapolating our results to a 40 ms integration time and 16 spectrally multiplexed channels yields an estimated sensitivity below 10 pW/nm in the H band. For comparison, the expected spectral flux density from Betelgeuse, collected with a 1 m telescope, is approximately 40 pW/nm according to the 2MASS catalogue.

## 4. CONCLUSION AND PERSPECTIVES

The present demonstration represents a first step toward broadband, spectrally multiplexed heterodyne interferometry. Although implemented here with a single spectral channel in the H band, the proposed architecture naturally extends to multi-channel operation and can be further expanded toward the K band by exploiting the broad bandwidth of mode-locked erbium-doped fiber lasers or by subsequent spectral broadening in nonlinear fibers. Importantly, commercially available broadband components, such as high-speed extended-InGaAs balanced detectors and 50/50 fiber couplers, are already compatible with this approach, providing a clear technological pathway toward large-scale spectral multiplexing and substantially enhanced sensitivity for applications in protoplanetary disks and evolved stellar environments.

## ACKNOWLEDGEMENTS

This work is supported by a grant from the French government, managed by the Agence Nationale de la Recherche under the Investissements d'Avenir UCAJEDI project, reference no. ANR-15-IDEX-01. We acknowledge the mechanical team members of the INPHYNI Sandra Bosio and Florian Zumbo for their support.